\documentclass[%
amsmath,amssymb,
aps,
]{revtex4-2}

\usepackage{graphicx}
\usepackage{dcolumn}
\usepackage{bm}
\usepackage{xspace}
\usepackage{xcolor}
\usepackage{definitions}

\usepackage[mathlines]{lineno}

\begin{document}

\preprint{APS/123-QED}


\title{Wave-like behaviour in (0,1) binary sequences}

\author{Enrique Canessa}
\affiliation{The Abdus Salam International Centre for Theoretical Physics, ICTP, Trieste 34151, Italy}

\date{\today}

\begin{abstract}
This work presents a comprehensive study of the properties of finite (0,1) binary systems from 
the mathematical viewpoint of quantum theory where a complex wavefunction $\psi_{n}$ is considered 
as an analogous probability measure. This is a quantum-inspired extension of the GenomeBits model 
to characterise observed genome sequences, where $\psi_{n}$ is related to an alternating (0,1) binary 
series having independent distributed terms. The real and imaginary spectrum of $\psi_{n}$ {\emph vs.} 
the nucleotide base positions display characteristic features of sound waves.
\end{abstract}

\keywords{binary systems, genome sequences, quantum theory, random numbers, sound waves}

\maketitle


\section{\label{sec:intro}Introduction}

The derivation of analogous relationships between different disciplines has become increasingly 
popular in recent years. This type of approach has been implemented in a wide range of applications, 
as for example in the domains of physics and finance. These efforts have pursue analogies between 
stochastic models commonly used in the statistical physics of complex systems and stock market 
dynamics~\cite{bib-1a}. The thermodynamic interpretation of multifractality was established 
in~\cite{bib-1b}, and an expression for an analogous specific heat in these systems was derived 
in~\cite{bib-1a}.

Another particularly interesting example is the 3D deformation of a compressible 
filament which has been modelled as the oscillation of a relativistic non-linear pendulum, where the 
compression modulus relates to the relativistic particle's rest mass, and the bending modulus 
mimics the speed of light~\cite{bib-1c}. A mathematical approach for the quantum information 
representation of biosystems has been introduced in~\cite{bib-1d}. In this study, biological functions 
such as psychological functions and epigenetic mutation are modelled in analogy with the physics 
of open quantum systems. In~\cite{bib-1e}, it has been argued how analogous Hawking radiation-like 
phenomena may arise in chaotic systems with exponential sensitivity to initial conditions (butterfly 
effect). There exists also suggestive parallels between various aspects of number theory and physics
phenomena~\cite{bib-1f}. 

Such parallel analyses have been useful to describe the rich complexity of diverse dynamical systems in 
terms of well-known physics phenomena and, therefore, best characterise the peculiarities of their behaviour.
Motivated by these analogous representations, we apply in this work quantum formalism outside of physics 
to derive properties of binary sequences containing 0 and 1 distinct outcomes from a new perspective. 

The purpose of this work is to introduce a wave-like function for complete genome sequences 
of pathogens represented by an alternating binary series having independently distributed terms associated 
with (0,1) binary indicators for the nucleotide bases. This quantum-based mathematical description is an 
extension of our previous GenomeBits model~\cite{bib-1g,bib-1h}. It can reveal further unique imprints 
of the genome dynamics at the level of nucleotide ordering for different systems (or genome mutations) 
following experimental measures over $N$ intervals. We compare results with random binary sequences.

Following~Ref.\cite{bib-1d}, the present approach is treated from the viewpoint of quantum theory as a 
measurement theory and not as being the atomic-level modelling of real quantum physical processes. The 
complex wavefunction is seen as a mathematical description of an isolated, analogous quantum system. 
The real and imaginary parts of the longitudinal wavefunction {\emph vs.} the nucleotide 
bases display analogous features of sound waves.

\section{\label{sec:formulae}Binary sequences formulae}

Let us consider the quantitative GenomeBits method for the examination of distinctive patterns of complete genome
data~\cite{bib-1g,bib-1h}. It consists of a certain type of finite alternating series having terms converted to 
(0,1) binary values for the nucleotide bases $\alpha = A,C,T,G$ as observed along the reported genome sequences, 
with (A)denine, (C)ytosine, (G)uanine and (T)hymine --or (U)racil RNA genome for single 
strand. In other words, the GenomeBits approach defines the sum relation 
\begin{equation}\label{eq:map}
     \Phi(X_{\alpha,k}) = \phi(X_{\alpha,1}) + \cdots + \phi(X_{\alpha,k}) = \sum_{j=1}^k \phi(X_{\alpha,j})=
                   \sum_{j=1}^k (-1)^{j-1}X_{\alpha,j} \; ,
\end{equation}
where the individual values $X_{j} = 0$ or $X_{j} = 1$ are associated according to their position $k$ along the 
genome sequences of length $N$, satisfying the following relation
\begin{equation}
     X_{\alpha,k} = | \Phi(X_{\alpha,k}) - \Phi(X_{\alpha,k-1}) | \; .  
\end{equation}
The arithmetic progression carries positive and negative signs $(-1)^{j-1}$ and a finite
non-zero first moment of the independently distributed variables $X_{\alpha,j}$. By default,
plus and minus signs are chosen sequentially starting with $+1$ at $j=1$. 

The mapping of~Eq.(\ref{eq:map}) into four binary projections of the $\alpha$-sequences follows 
the three-base periodicity characteristic of protein-coding DNA sequences studies in Ref.~\cite{bib-2a}.
Analysing genomics sequencing via this class of finite alternating sums allows to extract
unique features at each base. From the view of statistics, such series are equivalent to a 
discrete-valued time series for the statistical characterisation of random data sets~\cite{bib-1a}.
In the following, however, the above GenomeBits relation is to be considered as a resulting wave created 
by a certain superposition of (one or more discretized) wavefunctions, {\em i.e.}, 
$\phi(X_{\alpha,N}) \rightleftharpoons \psi_{n}(X_{\alpha,N})$ in some medium. In general wavefunctions 
are complex functions and the displacement of this wave is to be a function of base position $k$.

Let us then consider the polar form
\begin{eqnarray}\label{eq:psi}
\psi_{n}(X_{\alpha,k}) & \equiv & A\; (-1)^{k-1} | \Phi(X_{\alpha,k}) - \Phi(X_{\alpha,k-1}) |       
                          \; \exp\left\{\frac{n\pi i}{\lambda_{_{N}}}\Phi(X_{\alpha,k})\right\} \; ,  \nonumber \\
                           &    =    &   A\; (-1)^{k-1}X_{\alpha,k}       
                          \; \exp\left\{\frac{n\pi i}{\lambda_{_{N}}} \sum_{j=1}^k (-1)^{j-1}X_{\alpha,j}\right\} \; , 
\end{eqnarray}
with $A$ a real constant and $n = 1,2 \cdots$ 

The normalisation condition via the complex conjugate 
$\sum_{k=1}^N \psi_{n}(X_{\alpha,k})\psi_{n}^{*}(X_{\alpha,k}) = \sum_{k=1}^N |\psi_{n}(X_{\alpha,k})|^{2} = 1$ 
implies
\begin{equation}
A^{2}\sum_{k=1}^N |(-1)^{k-1}X_{\alpha,k}|^{2} = A^{2}\sum_{k=1}^N X_{\alpha,k}^{2} = 1 \; .
\end{equation}
Since $X_{\alpha,k}$ may take (0,1) values only, one then gets the amplitude
\begin{equation}\label{eq:a}
A = \frac{1}{\pm\sqrt{N_{_{+1}}}} \; ,
\end{equation}
where $N_{_{+1}}$ is the total number of 1's found in the complete $N = N_{_{0}} + N_{_{+1}}$ 
sequences for each species $\alpha$. 

To simplify calculations, let us choose
\begin{equation}
\lambda_{_{N}} \equiv \sum_{j=1}^N \phi(X_{\alpha,j}) = \sum_{j=1}^N (-1)^{j-1}X_{\alpha,j} = \Phi(X_{\alpha,N})\; ,
\end{equation}
which corresponds to the maximum real value for the alternating sum of binary sequences. From~Eq.(\ref{eq:psi}) 
and Euler's identity, it is interesting to note that for $k \rightarrow N$ the Cartesian form 
\begin{equation}\label{eq:psiN}
\psi_{n}(X_{\alpha,N}) = \left(\frac{1}{\pm\sqrt{N_{_{+1}}}}\right) (-1)^{N-1} X_{\alpha,N} \; [\cos(n\pi) + i \;\sin(n\pi)] 
                              =  \frac{(-1)^{n}}{\pm\sqrt{N_{_{+1}}}}\; \phi(X_{\alpha,N}) \; .
\end{equation}
oscillates and decreases for large numbers of 1's and $\forall n \geq 1$. Hence, the GenomeBits~Eq.(\ref{eq:map}) 
for complete genome sequences can be seen as a resulting wave in steady state created by certain non-zero 
complex wavefunctions $\psi$ that change with $k \ne 0$ and have same maximum density probability $1/N_{_{+1}}$ 
at each $n$.

\section{\label{sec:discussion}Results and discussion}

\begin{figure}[htb]
\includegraphics[width=0.6\textwidth]{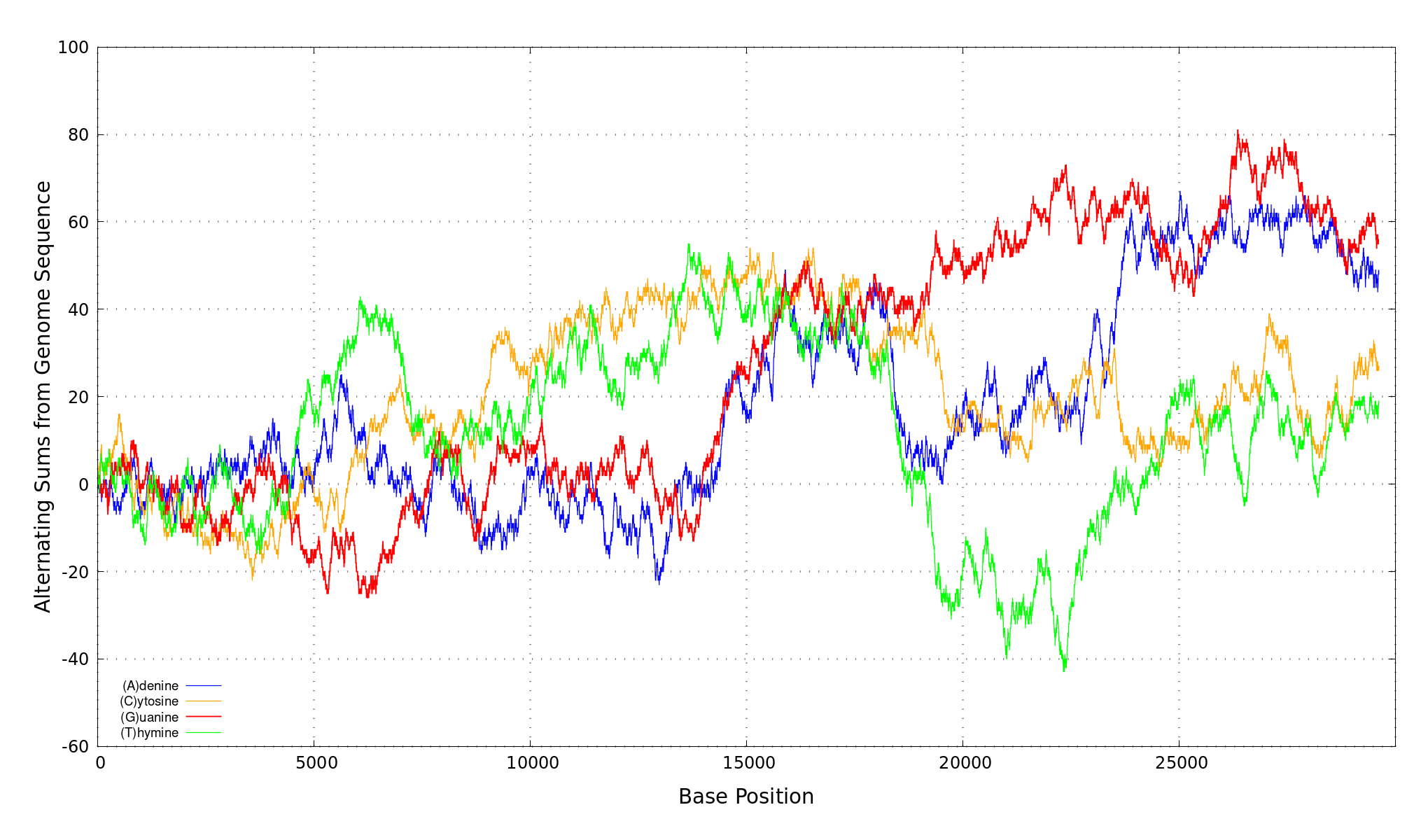}
\caption{\label{fig:fig1} 
Binary sequence sum series of~Eq.(\ref{eq:map}) for representative nucleotide bases from the genome sequence 
of coronavirus (Omicron variant, GISAID EPI\_ISL\_11901306~\cite{bib-3a}).
}
\end{figure}

In this section we discuss some examples addressed to show the applicability of the above results. 
The examples include cases of wavefunctions with random (0,1) sequences with either different densities 
of random 0's and 1's --namely, 70\% number of zeros and 30\% of ones, as well as a wavefunction 
for a representative full-length GenomeBits sequence of coronavirus (Omicron variant)~\cite{bib-1g,bib-1h}. 
Illustrative results derived for the sums in~Eq.(\ref{eq:map}) for nucleotides of this strand of complete 
genome sequence are shown in Fig.~\ref{fig:fig1}.

The GISAID accession number for the full-length genome sequencing shown in Fig.~\ref{fig:fig1} is EPI\_ISL\_11901306 
dated 2022-04-04 from Monza, Italy~\cite{bib-3a}, with $N=29613$ base pairs. It is interesting to note that for 
$\alpha = (A)$ the total numbers of zeros is 20770 (0.70\%) and the total number of ones is 8843 (0.30\%). 
For $\alpha = (C)$ these correspond to 24199 (0.82\%) and 5414 (0.18\%), respectively. For $\alpha = (G)$ one finds 
23809 (0.80\%) and 5804 (0.20\%) and, similarly, for $\alpha = (T)$ the values become 20061 (0.68\%) and 9552 (0.32\%). 
There are regions in this figure where these binary projections reveal some distinctive oscillatory patterns and 
interesting unique imprints of the intrinsic gene organisation at the level of nucleotides.

Remarkably, the signals in Fig.~\ref{fig:fig1} are essentially similar in behaviour. The positive and negative terms 
in the sums of the discrete (0,1) values partly cancel out, allowing the series to diverge from zero rapidly and to 
become a non-Cauchy sequence type. As shown next, these statistical representations are powerful to targeting the 
proposed complex-valued wavefunctions of~Eq.(\ref{eq:psi}).

\begin{figure}[!ht]
\includegraphics[width=0.7\textwidth]{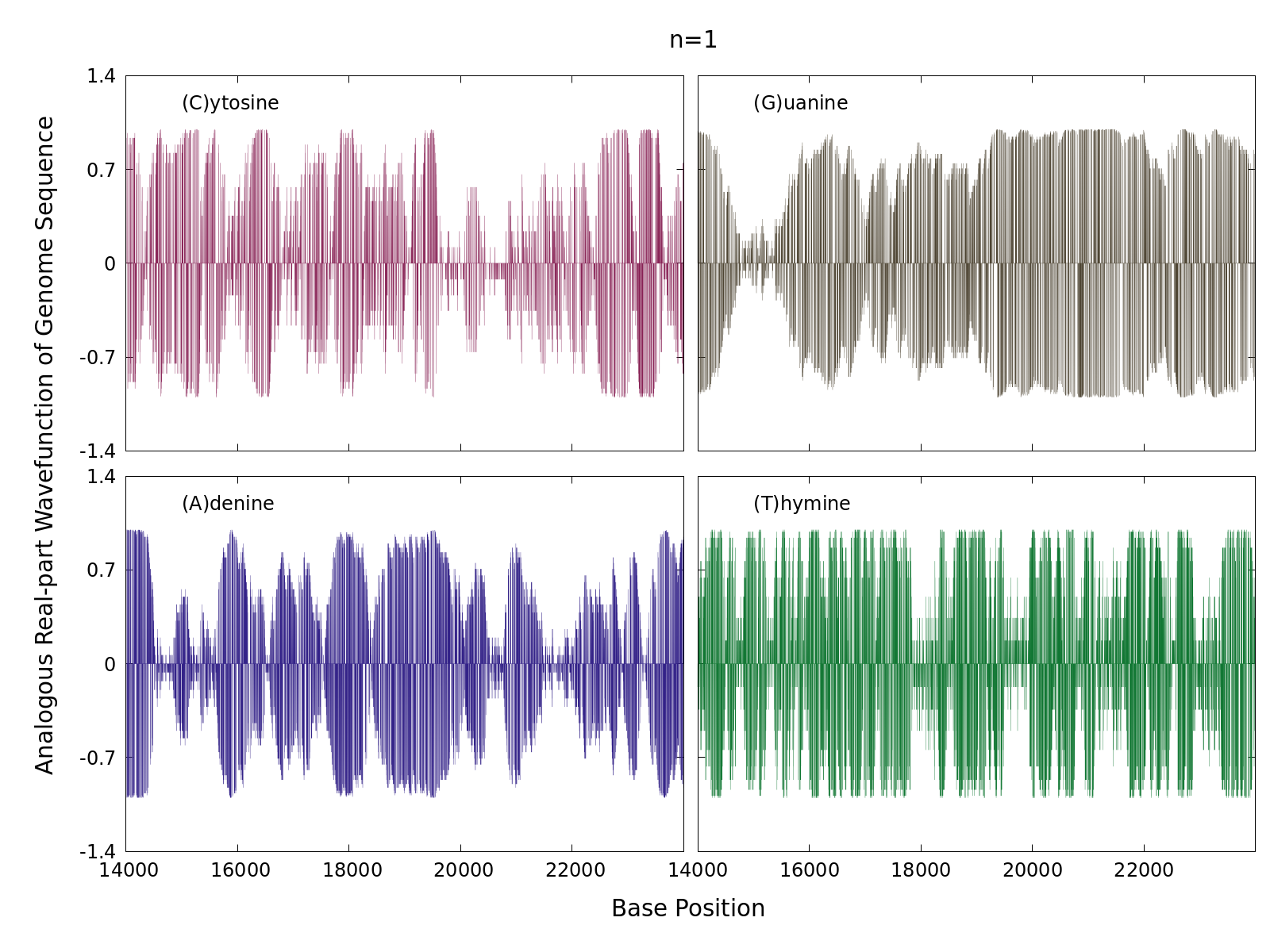}
\caption{\label{fig:fig2} 
Real parts of the wavefunctions in~Eq.(\ref{eq:psi}) for $n=1$ for each sum series displayed in Fig.~\ref{fig:fig1}.
}
\end{figure}

\begin{figure}[!ht]
\includegraphics[width=0.7\textwidth]{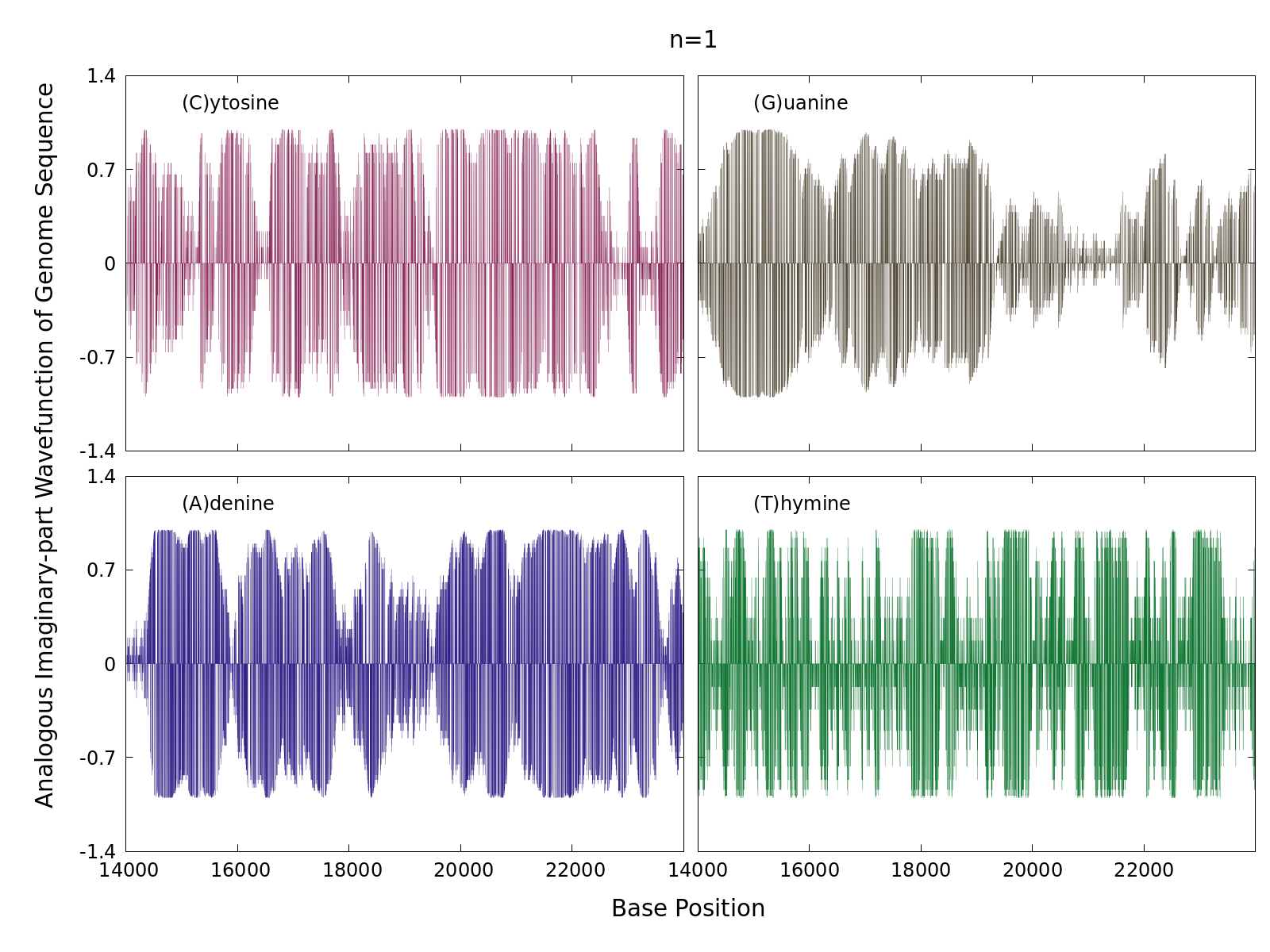}
\caption{\label{fig:fig3} 
Imaginary parts of the wavefunctions in~Eq.(\ref{eq:psi}) for $n=1$ for each sum series displayed in Fig.~\ref{fig:fig1}.
}
\end{figure}

The corresponding real and imaginary parts of $\psi_{1}(X_{\alpha,k})$ at a first analogous bound state are 
depicted in Figs.~\ref{fig:fig2} and~\ref{fig:fig3} for the sum series of Fig.~\ref{fig:fig1} along the base 
positions $14000 \le k \le 24000$ and normalized by $A$. This wavefunction is here understood as the mathematical 
description of an analogous ground state of an isolated quantum system. The real and imaginary parts may in turn 
resemble analogous oscillatory acoustic waves in all examples considered. Such sequences share zero values in 
$\mathbb{R}$ when $X_{\alpha,k}=0$. Individually all the wavefunctions for a given $n$ exhibit a set of 
zeros in their real and imaginary parts along the complete real $1 \le k \le N$ axis. 

These results are a consequence of factorising the full wavefunction as the product of two different expressions
{\emph i.e.}, a linear function proportional to the (0,1) binary sequences and a complex exponential form. Since 
this wavefunction is discretized, one may think of it as a standing wave with zero nodes. From the numerical point 
of view, these wavefunctions are here not quantum states of interest with a definite total physics energy, but rather 
an analogous mathematical construction. At any other point different from zero, the wavefunction with oscillatory 
behaviour for genome sequences is obtained. This is a remarkable result based on a simple algorithm of finite 
alternating sum series having independently distributed terms associated with just two indicators (0,1) for 
each of the nucleotide bases $A,C,T$ or $G$. In this light, it is worthy to note that previous algorithmic 
conversions of single-protein DNA sequence data to oscillatory sound waves has been reported in~Ref.\cite{bib-3c}, 
in which each DNA base is represented by specific musical notes. 

\begin{figure}[!ht]
\includegraphics[width=0.4\textwidth]{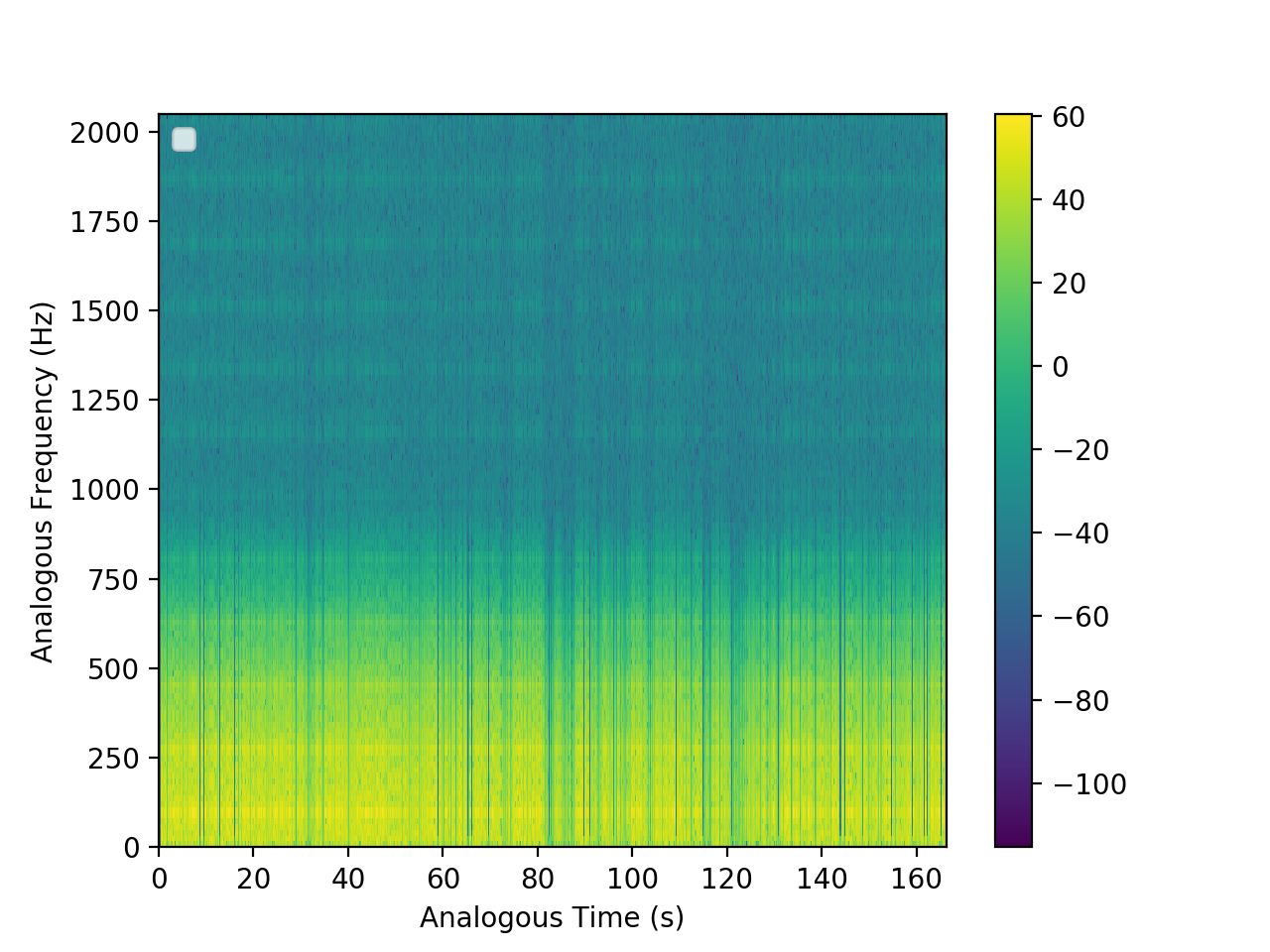}
\caption{\label{fig:fig4}
Example of analogous time-frequency representation of the strand of nucleotide bases (A)denine
produced from a wav audio file generated via the present GenomeBits wavefunction. The latter 
can be downloaded from GitHub~\cite{bib-3f}.
}
\end{figure}

The wav audio files generated via the present GenomeBits wavefunctions, by transforming the occurrence of 
nucleotides of the same type along the genome sequences can be downloaded from GitHub~\cite{bib-3f}. 
An analogous time-frequency representation of the strand of genome data, showing homogeneous peak signals 
over time, is shown in Fig.~\ref{fig:fig4} in the form of spectrogram. Each data point in the audio curves 
has been fitted to a gaussian function to correlate a continuous audio spectrum in the analysis of our 
waveform reconstruction. In these calculations, we use similar parameters as in~Ref.\cite{bib-3d}:
sample rate: 4096, precision: 16-bit, duration: 2:46.28 min for 681097 samples: file size: 1.36M: bit rate: 65.5k and
sample encoding: 16-bit in 1 channel. This visual method could also help to identify and quantify future virus mutations. 
In passing, we mention that gravitational waves from a binary black hole merger have been recently sonificated 
in~Ref.\cite{bib-3d}, and music from fractal noise have been investigated in~Ref.\cite{bib-3e}.

The present novel method of binary projections introduced here has drawn inspiration from quantum theory.
The main point is to uncover distinctive patterns out of some intrinsic organisation embedded in the 
sequences, including acoustic-like waves. Our work differs from other formalism in that sequence variations 
are characterised by simply associating (0,1) indicators for each nucleotide bases (A), (C), (G) and (T) 
separately.

\begin{figure}[!ht]
\includegraphics[width=0.7\textwidth]{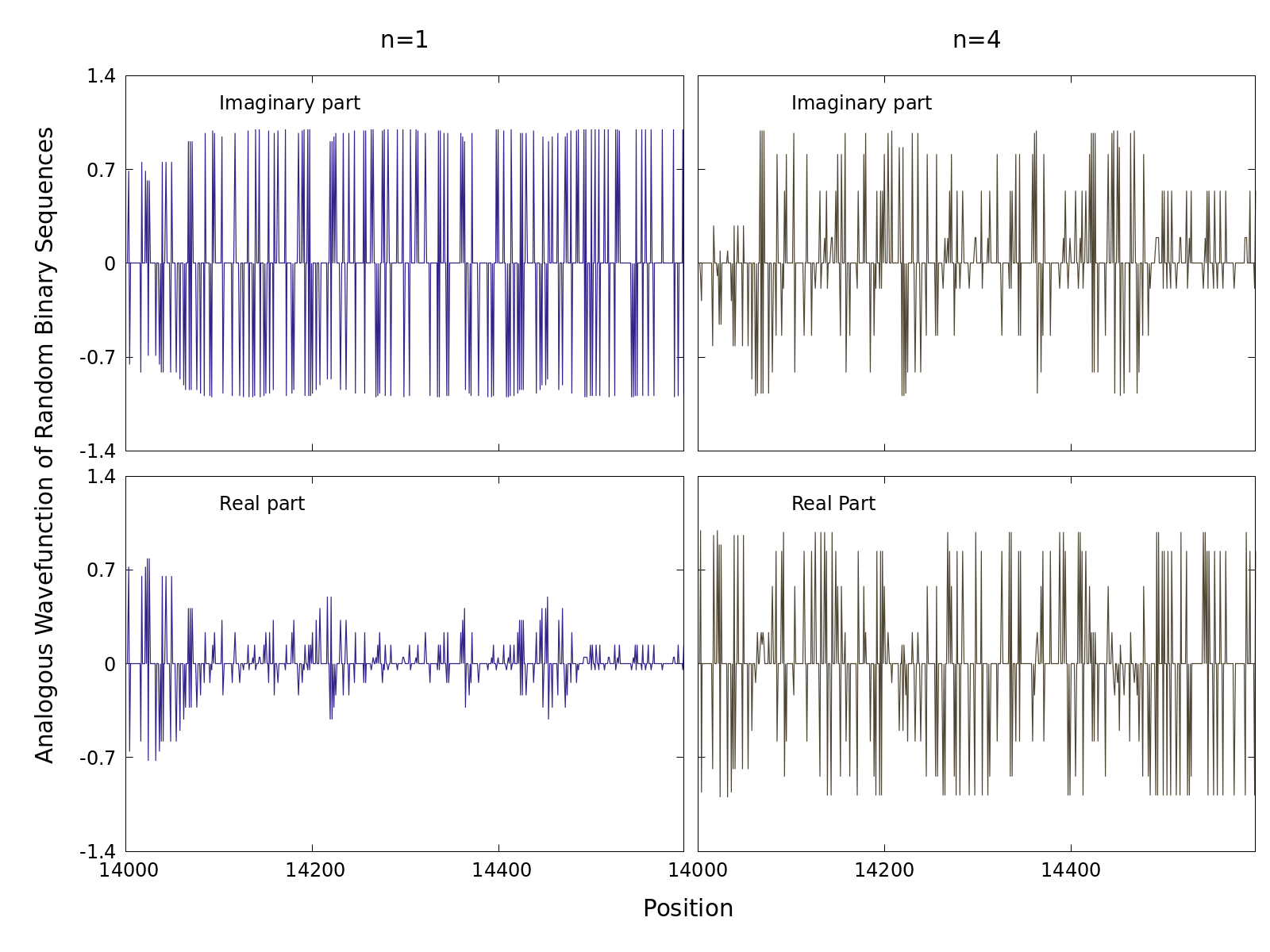}
\caption{\label{fig:fig5} 
Real and Imaginary parts of the wavefunctions in~Eq.(\ref{eq:psi}) for binary random numbers with 
different frequencies of 0's (70\%) and 1's (30\%) and two different $n$'s.
}
\end{figure}

Inspired by the above findings in the genome sequences of coronavirus, we include in Fig.~\ref{fig:fig5} 
examples addressed to show that, in general, there are common patterns in the behaviour around the zeros 
of the real and complex parts of the wavefunction $\psi_{1}(X_{\alpha,k})$ for random (0,1) sequences 
with 70\% of number of zeros and 30\% of ones. For different sequences randomly generated in arbitrary 
form and for two different values of $n$, it follows that the maxima and minima in the curves differ in 
shape. For $n=1$ the profile of the imaginary part of $\psi$ presents an analogous "white noise" 
type of behaviour, which carries the least correlations. As for Fig.~\ref{fig:fig3}, the occurrence 
of a particular bit is essentially independent of previous or future values of the binary 
expansion~\cite{bib-3b}. 

Closing this discussion, it seems sensible to interpret our quantum inspired~Eq.(\ref{eq:psi}) as
an inherent mathematical method to study oscillations in binary systems although its nature may  
appear abstract. We stress that classical mathematics still plays a crucial role to describe the 
rich complexity of biological systems and peculiarities of their behaviour~\cite{bib-1d}.

\section{\label{sec:summary}Summary}

Starting with an extension of the alternating sums of genome sequences as in Eq.(\ref{eq:map})
(which may be seen as a particular case of polynomial associated to some immanant of a matrix 
for DNA graph~\cite{bib-4a}), we presented a new quantitative test for comparing natural binary 
numbers sequences through the complex wavefunction of~Eq.(\ref{eq:psi}). Its analogous 
cumulative probability over all possible base positions is unity. This measure is aimed at finding 
out a characterisation and understanding of sequence variation from a comparative investigation 
across multiple binary systems including genome sequences and random numbers. Although the present 
analogy is consistent from a formal mathematical perspective, one aspect that can be criticised 
is the fact that we cannot make yet some kind of physical association to quantize energy in levels, 
to derive a system momentum, {\emph etc}. The justification for this is that we are dealing with 
discretized (0,1) sequences in one dimension.

By construction, the wavefunction~Eq.(\ref{eq:psi}) at $n$ wear properties that are not different 
from those of other analogous states $n+1$ as can be deduced through~Eq.(\ref{eq:psiN}). For finite 
systems, it makes sense to fix the amplitude $A$ by the normalisation of an analogous total probability 
as done in~Eq.(\ref{eq:a}). Under base translation by a spacing $k_{a} > k$, the wavefunction 
is not symmetric and its modulus satisfies
$|\psi_{n}(X_{\alpha,k+k_{a}})|^{2} = |X_{\alpha,k+k_{a}}|^{2} \ne |X_{\alpha,k}|^{2}$.
Such behaviour is characteristic of the multi-mode spectrum of aperiodic lattices.  Additionally, 
the observations reported here may share a common quantum-like origin with the method introduced
in~\cite{bib-1d}.

The examples worked out in this paper have been chosen as another test for the mathematical conversion of binary
sequences to isolate acoustic-like waves  by appropriate formulas. Introducing~Eq.(\ref{eq:psi}) as an extension 
of~Eq.(\ref{eq:map}), we developed a simple method to perform wave calculations where inputs of zeros 
and ones are sufficient. This opens the possibility to reverse the calculations and derive a characteristic precursory 
alternating sum from a given recorded audio file as input. We believe this novel aspect of our algorithm could 
stimulate further investigations toward slicing the sounds of nature.

\section*{\label{sec:declaration}Conflict of interest}

The author declares no conflict of interest.

\end{document}